\documentclass[preprint,showpacs,amsmath,amssymb,superscriptaddress]{revtex4}
\usepackage[dvips]{graphicx}
\usepackage{dcolumn}
\usepackage{bm}
\usepackage{color}
\usepackage{ulem}

\newcommand{\bs}   {\boldsymbol}  
\newcommand{\mb}   {\mathbf}      
\newcommand{\mr}   {\mathrm}      
      
\newcommand{\mcal} {\mathcal}     
\newcommand{\imag} {\mathrm{i}}   
   
\newcommand{\e}    {\mathrm{e}}   

\newcommand{\w}    {\omega}

\begin{document}


\title{Excitonic Bose-Einstein condensation above room temperature}

\author{K.~Seki}
\affiliation{Department of Physics, Chiba University, Inage-ku, Chiba 263-8522, Japan}
\author{Y.~Wakisaka}
\affiliation{Department of Physics and Department of Complex Science and Engineering, University of Tokyo, Kashiwa, Chiba 277-8561, Japan}
\author{T.~Kaneko}
\affiliation{Department of Physics, Chiba University, Inage-ku, Chiba 263-8522, Japan}
\author{T.~Toriyama}
\affiliation{Department of Physics, Chiba University, Inage-ku, Chiba 263-8522, Japan}
\author{T.~Konishi}
\affiliation{Department of Nanomaterial Science, Chiba University, Inage-ku, Chiba 263-8522, Japan}
\author{T.~Sudayama}
\affiliation{Department of Physics and Department of Complex Science and Engineering, University of Tokyo, Kashiwa, Chiba 277-8561, Japan}
\author{N.~L.~Saini}
\affiliation{Department of Physics, Sapienza University of Rome, 00185 Rome, Italy}
\author{M.~Arita}
\affiliation{Hiroshima Synchrotron Radiation Center, Hiroshima University, Higashi-Hiroshima 739-0046, Japan}
\author{H.~Namatame}
\affiliation{Hiroshima Synchrotron Radiation Center, Hiroshima University, Higashi-Hiroshima 739-0046, Japan}
\author{M.~Taniguchi}
\affiliation{Hiroshima Synchrotron Radiation Center, Hiroshima University, Higashi-Hiroshima 739-0046, Japan}
\affiliation{Graduate School of Science, Hiroshima University, Higashi-Hiroshima 739-8526, Japan}
\author{N.~Katayama}
\affiliation{Department of Applied Physics, Nagoya University, Chikusa-ku, Nagoya 464-8601, Japan}
\author{M.~Nohara}
\affiliation{Department of Physics, Okayama University, Tsushima-naka, Okayama 700-8530, Japan}
\author{H.~Takagi}
\affiliation{Department of Physics, University of Tokyo, Bunkyo-ku, Tokyo 113-0033, Japan}
\author{T.~Mizokawa}
\affiliation{Department of Physics and Department of Complex Science and Engineering, University of Tokyo, Kashiwa, Chiba 277-8561, Japan}
\author{Y.~Ohta}
\affiliation{Department of Physics, Chiba University, Inage-ku, Chiba 263-8522, Japan}

\date{\today}

\begin{abstract}
We show that finite temperature variational cluster approximation (VCA) calculations 
on an extended Falicov-Kimball model can reproduce angle-resolved photoemission spectroscopy 
(ARPES) results on Ta$_2$NiSe$_5$ across a semiconductor-to-semiconductor structural phase
transition at 325 K. We demonstrate that the characteristic temperature dependence 
of the flat-top valence band observed by ARPES is reproduced by the VCA calculation 
on the realistic model for an excitonic insulator only when the strong excitonic fluctuation 
is taken into account. 
The present calculations indicate that Ta$_2$NiSe$_5$ falls in the Bose-Einstein condensation
regime of the excitonic insulator state.
\end{abstract}

\pacs{71.35.Lk, 71.20.-b, 67.85.Jk, 03.75.Nt}

\maketitle

\section{Introduction}

The diversity of Bose-Einstein condensation (BEC) with spin (and orbital) 
degrees of freedom is one of the main targets in the research of ultracold atom systems.
\cite{Wang2010,Lin2011,Ho2011}
On the other hand, in electronic systems with hole-like and electron-like Fermi pockets,
interaction between hole and electron in the Fermi pockets tends to induce charge density wave
(CDW) or spin density wave (SDW) with wave vector $Q$ which spans the hole-like 
and electron-like Fermi pockets in the Brillouin zone.
Such CDW or SDW transitions can be described as Peierls transitions 
due to Fermi surface nesting.
An alternative way to understand CDW or SDW transitions in the electronic systems
is the theory for excitonic insulators that was proposed nearly 
half a century ago. \cite{Mott1961, Jerome1967, Zittartz1967, Halperin1968}
Since the electron and hole concentrations are relatively low,
Coulomb interaction between holes and electrons in the Fermi pockets is 
weakly screened and stabilizes electron-hole bound states (excitons)
with wave vector $Q$.
In the Peierls picture and the excitonic insulator picture for metallic systems,
the CDW or SDW transitions are basically BCS type.
On the other hand, the theory of excitonic insulators can be extended to
semiconductors where excitons are formed due to Coulomb interaction between 
valence band holes and conduction band electrons and undergo BEC.
Very recently, BCS-BEC crossover of excitonic insulator has been examined
in more advanced calculations using such as Mott-Wannier-type exciton model 
with $T$-matrix approximation 
\cite{Bronold2006} and extended Falicov-Kimball model (EFKM).
\cite{Ihle2008,Phan2010,Seki2011}
If the normal state above the excitonic transition temperature is 
semimetallic namely the magnitude of the band gap $E_G < 0$, the excitonic transition is
well described in the BCS framework. On the other hand, 
the excitonic transition is expected to be a BEC of excitons
if the normal state is semiconducting or $E_G > 0$ as schematically
shown in Fig. 1(a).
In the BEC regime with $E_G > 0$, large amount of excitons 
are preformed above the excitonic transition temperature and
the excitonic transition corresponds to the BEC of excitons.   

The BEC transition temperature of excitons can be much higher 
than those of liquid He or cold atoms since the electron and 
hole masses are much smaller than atomic masses.  
Therefore, the BEC regime of the excitonic insulator can 
provide a unique opportunity to study ``high temperature'' BEC.
Real materials studied in relevance to excitonic insulator states 
are limited to several rare-earth or transition-metal 
calcogenides such as Tm(Se,Te) \cite{Neuenschwander1990, Wachter2004} 
and $1T$-TiSe$_2$. \cite{Cercellier2007}
The possible excitonic insulator transition in Tm(Se,Te) has been studied
using various transport measurements including electrical resistivity \cite{Neuenschwander1990}
and heat conductivity. \cite{Wachter2004} Although the increase of electrical resistivity
at the transition and the increase of heat conductivity at the low temperature are
assigned to the BEC and the superfluidity of excitons respectively, the excitonic 
insulator transition in Tm(Se,Te) is not established well due to the lack of
quantitative theory.
On the other hand, the CDW transition in semimetallic $1T$-TiSe$_2$,
which has hole-like (electron-like) Fermi pockets 
at the zone center (boundary), has been identified 
as a BCS-like excitonic insulator transition by angle-resolved 
photoemission spectroscopy (ARPES). \cite{Cercellier2007}
The BCS-like excitonic transition in $1T$-TiSe$_2$ is a semimetal-to-semimetal 
transition due to the small off-stoichiometry or the small difference between
the hole and electron densities. Therefore, it is rather difficult to
characterize the excitonic transition by means of transport measurement,
and the band dispersion observed by ARPES is the very direct evidence 
of the excitonic transition. The BCS-like excitonic transition
can be described by the mean-field theory and the ARPES result
can be compared with the theoretical calculation. \cite{Cercellier2007}
As for the BEC-like excitonic transition, it has been proposed that 
Ta$_2$NiSe$_5$ would have an excitonic insulator transition in the BEC regime
based on the ARPES results below the transition temperature. 
\cite{Wakisaka2009} The BEC-like excitonic transition is an insulator-insulator
transition, and it is rather difficult to identify the BEC of excitons 
using transport measurement. Similar to the BCS-like excitonic transition
in $1T$-TiSe$_2$, the band dispersion observed by ARPES can be one of the most
straightforward evidence of the excitonic transition. Namely,
the single particle excitation can be a finger print of the excitonic BEC 
just like the opening of superconducting gap in the BEC of Cooper pairs.  
However, the BEC regime of the excitonic insulator transition cannot be 
described by the mean-field theory and, therefore, a new theory beyond mean-field 
approximation should be developed and compared with the ARPES result 
in order to establish the excitonic insulator transition in the BEC regime.
In this sense, the BEC-like excitonic transition is not well established yet 
in any real materials including Ta$_2$NiSe$_5$.

As illustrated in Fig. 1(b),
Ta$_2$NiSe$_5$ has a quasi-one-dimensional (quasi-1D) structure 
where Ni and Ta atoms are arranged in one dimensional chains. 
\cite{Sunshine1985, DiSalvo1986}
Resistivity of Ta$_2$NiSe$_5$ is insulating below 500 K and
exhibits an anomaly around 325 K which is assigned as second-order 
or weak first-order structural phase transition without CDW.
The resistivity exhibits insulating behaviors both above and below
the transition temperature at 325 K. The magnetic susceptibility
shows a gradual drop around 325 K suggesting spin singlet formation.
The flattening and sharpening of the valence band top 
in insulating Ta$_2$NiSe$_5$ suggests that the phase transition at
325 K corresponds to an excitonic insulator transition in the BEC regime.
In addition, Ta$_2$NiSe$_5$ is unique in that the excitons are 
formed between the electrons and holes that are located in spatially 
separated chains: i.e., holes in the Ni chain and electrons 
in the two neighboring Ta chains in the quasi-1D structural units, 
which has an interesting similarity with the 2D bilayer electron 
systems of semiconductors where the electrons and holes are also 
spatially separated \cite{Eisenstein}.  
In the present work, we have developed a finite temperature variational 
cluster approximation (VCA) methods on EFKM 
for Ta$_2$NiSe$_5$ and have applied it to interpret temperature-dependent 
ARPES results of Ta$_2$NiSe$_5$ across the transition and 
to reveal the nature of the transition in this material.

\section{extended Falicov-Kimball model}

In order to include the electron correlation effects on the single-particle 
excitation spectra of Ta$_2$NiSe$_5$ across the excitonic insulator transition, 
we have employed VCA calculations on EFKM which is known as
a minimal lattice model to describe the excitonic insulator state.
\cite{Ihle2008,Phan2010,Seki2011}
The Ta 5$d$ and Ni 3$d$ orbital degeneracy at the conduction-band bottom and the valence-band top 
is removed due to the quasi-one-dimensional structure. \cite{Seki2011} Therefore,
in order to describe the excitonic condensation and spectral function of valence-band top, 
the Coulomb interaction between the valence-band hole and the conduction-band electron 
is enough and, consequently, the EFKM with non-degenerate and spinless 
conduction and valence bands is sufficient.
The Hamiltonian of the EFKM \cite{Ihle2008,Phan2010,Seki2011} is given as

\begin{eqnarray}\label{eq.ham}
{\cal H} &=& 
   - \sum_{\delta=x,y,z} t_c^{\delta} \sum_{\langle ij \rangle} (c_i^{\dag} c_j + \mathrm{H.c.}) + (D/2 - \mu) \sum_{i} n_{ic} \nonumber \\
&& - \sum_{\delta=x,y,z} t_f^{\delta} \sum_{\langle ij \rangle} (f_i^{\dag} f_j + \mathrm{H.c.}) + (-D/2 - \mu) \sum_{i} n_{if} \nonumber \\
&&+ U \sum_{i} n_{i c} n_{i f},
\end{eqnarray}

\noindent where $c_i$ ($c_i^\dag$) denotes the annihilation (creation) operator 
of an electron on the $c$-orbital (corresponding to the Ta 5$d$ conduction band) 
at site $i$ and $n_{ic} = c_{i}^\dag c_{i}$ is the particle number operator.
$t_c^{\delta}$ represents the spatially anisotropic hopping integral 
between neighboring sites parallel ($\delta=x$) and perpendicular ($\delta=y,z$) 
to the one-dimensional chain direction,
as it is realized in the quasi-one-dimensional material Ta$_2$NiSe$_5$.
These are the same for the $f$-orbital (corresponding to the Ni 3$d$-Se $4p$ valence band).
$D$ is the on-site energy split between the $c$- and $f$-orbitals and 
$U$ is the inter-orbital Coulomb repulsion between electrons.
Note that $U$ can be considered as inter-orbital electron-hole attraction 
and thus can drive the excitonic insulator state in this model at low temperature.
The effect of the electron-lattice interaction has been examined in Ref. \cite{Kaneko2013} 
and it has been found that the electron-lattice interaction plays a secondary role 
and helps the excitonic transition driven by the electron-hole interaction.
\cite{Kaneko2013}
The chemical potential $\mu$ is determined so as to maintain the average particle density at half filling.
The order parameter is given by

\begin{eqnarray}
2\Delta = U\langle c_i^{\dag}f_i + \mathrm{H.c.} \rangle. 
\end{eqnarray}

\noindent We use a parameter set of 
$t_c^x = -t_f^x = 0.4$eV, 
$t_{c}^{y}=-t_{f}^{y}=-t_c^{z}=t_f^{z}=0.024$eV, 
$D/2=0.44$eV, and 
$U=0.84$eV
so as to reproduce the single-particle excitation spectrum observed in ARPES 
for Ta$_2$NiSe$_5$ and the transition temperature at the same time.
The dependence on the cluster size of the calculated results was checked by changing the cluster size 
such as the one-dimensional 4-, 5-, 6-site (8-, 10-, and 12-orbital) clusters, where each site contains 
a pair of the $c$- and $f$-orbitals, and no significant cluster-size dependence was found.
Here, it should be noted that this parameter set corresponds to the semimetallic state
in the non-interacting ($U=0$) limit while it can describe the semiconducting
state due to the Hartree shift by $U$. \cite{Phan2010}

\section{Methods}

\subsection{Variational cluster approximation}

The VCA~\cite{Potthoff2003} with its finite temperature algorithm was employed to calculate 
the temperature dependence of the excitonic order parameter and the single-particle excitation spectra. 
Here, let us describe the formulation of the variational cluster approximation (VCA) at finite temperature 
used in this article.

First we briefly review the rigorous variational principle for the thermodynamic potential functional
derived by Luttinger and Ward~\cite{Luttinger1960} and revisited by Potthoff. \cite{Potthoff2003}
It has been shown by Potthoff~\cite{Potthoff2003} that the thermodynamic potential functional 
as a functional of the self-energy can be written as
\begin{equation}
\Omega[ \bs{\Sigma} ] = {\cal F} [ \bs{\Sigma} ] - \mr{Tr} \ln \left( -\bs{G}_{0}^{-1} + \bs{\Sigma} \right),
\end{equation} 
where ${\cal F} [\bs{\Sigma}]$ is the Legendre transform of the Luttinger-Ward functional $\Phi[\bs{G}]$,~\cite{Luttinger1960} 
$\bs{G}_{0}$ is the non-interacting Green's function, 
and the self-energy $\bs{\Sigma}$ is considered as a trial function for the variational calculation described below.
$\mr{Tr}$ represents the sum over fermionic Matsubara frequencies $\imag \w_\nu = (2\nu + 1 )\pi T$,
where $T$ is the temperature and $\nu$ is an integer,  and trace over the single-particle basis.
The stationarity condition 
\begin{equation}\label{eq.VP}
\delta \Omega[\mb{\Sigma}]/\delta \bs{\Sigma} = 0
\end{equation}
gives the Dyson equation and the functional at the stationary point gives 
the thermodynamic potential of the system.~\cite{Potthoff2003,Luttinger1960}

The self-energy functional theory (SFT)~\cite{Potthoff2003} provides a way to compute $\Omega$
by using the fact that the functional form of ${\cal F} [\bs{\Sigma}]$ depends only on 
the interaction terms of the Hamiltonian.
In SFT, 
the original lattice is divided into disconnected finite-size clusters and 
the reference system is introduced as their collection.
The clusters form a superlattice and have the same interaction term as the original system  
because the interaction term of the Hamiltonian 
of the extended Falicov-Kimball model [EFKM, Eq.~(1)]
is local in real space. 
Therefore the functional form ${\cal F}[\bs{\Sigma}]$ of the reference system is 
the same for that of the original system. 
The exact thermodynamic potential of the reference system is given by 
$\Omega' = {\cal F} [\bs{\Sigma}'] - \mr{Tr} \ln (-\bs{G}{'}_{0}^{-1} + \bs{\Sigma}')$, 
where $\bs{G}{'}_0$ and $\bs{\Sigma}'$ are the non-interacting Green's function and 
the exact self-energy of the reference system, respectively.
Then, by restricting the self-energy $\bs{\Sigma}$ to $\bs{\Sigma}'$, 
we can eliminate the functional ${\cal F} [\bs{\Sigma}']$ and obtain 
\begin{equation}\label{eq.Omega}
\Omega [\bs{\Sigma}'] = \Omega' - \mr{Tr} \ln \left( \bs{I}-\bs{V}\bs{G}{'} \right) 
\end{equation}
where $\bs{I}$ is the unit matrix, 
$\bs{V}  \equiv  \bs{G}{'}_{0}^{-1} - \bs{G}_{0}^{-1}$ represents the difference of 
the one-body terms between the original and reference systems,
and $\bs{G}{'}=(\bs{G}{'}_{0}^{-1} - \bs{\Sigma}')^{-1}$ is the exact Green's function of the reference system. 
The size of these matrices are $2L_{\mr{c}} \times 2L_{\mr{c}}$, where $L_{\mr{c}}$ 
is the number of sites within a disconnected finite-size cluster.

The Hamiltonian of the reference system is defined as 
\begin{eqnarray}
&{\cal H}'            &= {\cal H} + {\cal H}_{\mr{EI}}, \\
&{\cal H}_{\mr{EI}} &= \Delta' \sum_{i} \left(c_{i}^\dag f_{i}  + \mr{H.c.} \right),
\end{eqnarray}
where ${\cal H}$ is the Hamiltonian of the EFKM defined in Eq.~(1),
and  $\Delta'$ is a Weiss field for excitonic condensation which is treated as a variational parameter.
In other words, the self-energy $\bs{\Sigma}$ is varied through $\Delta'$ and thus the varitational principle Eq.(\ref{eq.VP}) is 
practically given by
$\partial \Omega/\partial \Delta'= 0$. 
Note that the solution with $\Delta' \not= 0$ corresponds to the spontaneous EI state.
Then we solve the eigenvalue problem ${\cal H}'|\Psi_n \rangle = E_n |\Psi_n\rangle $ for all excited states
of a finite-size cluster and calculate the trial single-particle Green's function. 
The Green's function matrix in Eq.~(\ref{eq.Omega}) is defined as  
\begin{eqnarray}
\bs{G}{'} (\w)=
\left( \begin{array}{cc}
\bs{G}{'}^{cc} (\w) & \bs{G}{'}^{cf} (\w) \\
\bs{G}{'}^{fc} (\w) & \bs{G}{'}^{ff} (\w)
\end{array} \right),
\end{eqnarray}
with the matrix elements of 
\begin{equation}
G{'}^{cf}_{ij} (\imag \w_\nu) = \e^{\beta \Omega'} \sum_{r,s}
\left(\e^{-\beta E_r} + \e^{-\beta E_s}\right)
\frac{\langle \Psi_r |c_i^\dag |\Psi_s  \rangle 
      \langle \Psi_s |f_j     |\Psi_r  \rangle}{\imag \w_\nu - (E_r - E_s)},
\end{equation}
where $|\Psi_r\rangle$ is an eigenstate of $\mcal{H}'$ and $E_r$ is the corresponding eigenvalue.

\subsection{Cluster perturbation theory}

The single-particle excitation spectra are calculated 
by the cluster perturbation theory (CPT).~\cite{Senechal2000} 
In the CPT, the single-particle Green's function is given as
\begin{equation}
\mcal{G}^{\alpha} (\mb{k},\w) = 
\frac{1}{L_{\mr{c}}}  
\sum_{i,j}^{L_{\mr{c}}} 
G^{\alpha}_{\mr{CPT},ij} (\mb{k},\w) 
\e^{-\imag \mb{k} \cdot (\mb{r}_i - \mb{r}_j)},
\end{equation}
where
$\alpha$ ($=c,f$) denotes the orbital index,  
$\mb{r}_{i}$ is the position of the $i$-th site within a cluster, and 
\begin{equation}
\bs{G}_{\mr{CPT}}^{\alpha}(\mb{k},\w) = 
\left( \bs{G}'(\w) [\bs{I}-\bs{V}(\mb{k})\bs{G}'(\w)]^{-1} \right)^{\alpha \alpha}
\end{equation}
is the CPT Green's function \cite{Senechal2000}.
Note that the wave vector $\mb{k}$ can take arbitrary values
in the Brillouin zone.
The CPT is exact both in the non-interacting $(U=0)$ and atomic 
$(t=0)$ limits, and is expected to work well in strongly interacting 
regime since it is derived originally from the strong-coupling expansion 
for the single-particle Green's function.
In the weak- to intermediate-coupling regime, 
the CPT approximation is generally improved with increasing the cluster size. \cite{Senechal2000} 
Thus larger clusters are in principle required to obtain more accurate results. 
We have checked calculations of single-particle excitation spectra 
as well as the excitonic order parameters for 
one-dimensional clusters of the size $L_{\mr{c}} = 4,5,$ and $6$ (8-,10-, and 12-orbital),
where each site contains a pair of the $c$- and $f$-orbitals, 
and have found no significant cluster-size dependence.
In the main text, calculated results for the $L_{\mr{c}}=5$ (10-orbital) cluster are shown,
thus the effects of statical and dynamical electron correlation within the cluster size 
are taken into account. 

\subsection{Angle-resolved photoemission spectroscopy}
 
The ARPES data compared to the calculated results were taken at beamline 9A, 
Hiroshima Synchrotron Radiation Center (HSRC) using a Scienta R4000 analyzer 
with an angular resolution of $\sim 0.3^\circ$ and energy resolution $\sim 26$ meV
using circularly polarized light of photon energy $h\nu=$23 eV.
The ARPES data with higher quality than those reported in ref. 15 \cite{Wakisaka2009} 
cover the temperature range from 40 K to 360 K including the transition temperature $\sim$ 325 K.
The ARPES data partially reported in ref. 20 \cite{Wakisaka2012} 
were reanalyzed considering the calculated results and are presented.
ARPES spectra are obtained as $\rho_k(E)$ where $E$ and $k$ are electron energy and wave vector.
In the second derivative plot, $-d^2\rho_k(E)/dE^2$ is smoothed and is
plotted as a function of electron energy and wave vector.

\section{Results and discussion}

\subsection{Transport properties}

Figure 2 shows the electrical resistivity $\rho$ and $-T^2 dln\rho/dT$ 
as a function of temperature $T$ as well as the magnetic susceptibility 
$M/H$ of Ta$_2$NiSe$_5$.
The electrical resistivity $\rho$ and $-T^2 dln\rho/dT$ of Ta$_2$NiSe$_5$ 
exhibit an anomaly at 325 K which corresponds to the excitonic insulator transition. 
The resistivity exhibits insulating behaviors both above and below 
the transition. If temperature dependence of resistivity is 
given by a simple activation type function $e^{E_G/k_BT}$ 
with band gap of $E_G$, $\rho$ and $-T^2 dln\rho/dT$ provides 
magnitude of $E_G/k_B$. $E_G$ is estimated to be $\sim$ 1000 K
which is consistent with the magnitude of the gap observed by ARPES.
The magnetic susceptibility shows a gradual drop around 325 K
which can be assigned to the gradual increase of the band gap 
due to the excitonic coupling between the valence band hole and 
the conduction band electron.

\subsection{Temperature dependence of ARPES spectra}

Figure 3 shows temperature dependence of ARPES spectra around the $\Gamma$ point.
In order to capture the temperature evolution of the band dispersions,
the valence-band-peak positions evaluated from the energy distribution curves
(which are indicated by the dots with the error bars in Fig. 3) were fitted 
to the model function $\epsilon_v(k)$ for the excitonic insulator
band dispersion (shown by the thick solid curves in Fig. 3).
The model function $\epsilon_v(k)$ for the excitonic insulator
band dispersion is obtained as below.
\begin{align} \epsilon^0_v(k)&=\epsilon_1+\sqrt{(\epsilon_2/2)^2+(2t_{\mathrm{Ni-Se}}\cos^2ka/2)^2} \\
  \epsilon^0_c(k)&=\epsilon_3-2t_{\mathrm{Ta}}\cos ka \\
  \epsilon_v(k)&=\frac{\epsilon^0_v(k)+\epsilon^0_c(k)}{2}-\sqrt{\left(\frac{\epsilon^0_v(k)-\epsilon^0_c(k)}{2}\right)^2+\Delta^2}.
\end{align}
Here, $\epsilon^0_v(k)$ and $\epsilon^0_c(k)$ are the bare valence and 
conduction bands derived by a tight-binding model 
free from excitonic interaction. In the tight-binding model,
the Ni-Se and Ta chains are considered to form one-dimensional band dispersion 
along the chain direction only through their nearest neighbor 
transfer integrals $t_{\mathrm{Ni-Se}}$ and $t_{\mathrm{Ta}}$ 
without inter-chain hybridization.
Assuming linear temperature dependence of the chemical potential, 
the model function $\epsilon_v(k)$ is fitted to the valence-band-peak positions
to optimize $t_{\mathrm{Ni-Se}}$ and the excitonic insulator order parameter $\Delta$ 
as adjustable parameters. 
The fitting results are shown by the thick solid curves in Fig. 3,
indicating that the fits are reasonably good to extract 
$\Delta$ as a function of temperature. 
The bare valence-band dispersion $\epsilon^0_v(k)$ without the excitonic coupling
is indicated by the thin solid curves in Fig. 3.

\subsection{Comparison between VCA calculations and ARPES results}

Overall valence-band dispersions observed by ARPES are consistent with the tight-binding 
or LDA band structure calculations \cite{Canadell1987,Kaneko2013}. 
However, there is a prominent discrepancy near the top of the valence band 
located at the $\Gamma$ point where characteristically flat band 
dispersion is observed as shown in Fig. 4(a).
This flat band dispersion reminds the experimental ARPES result of 
$1T$-TiSe$_2$ which is one of the candidate materials 
for showing excitonic insulator transition. \cite{Cercellier2007,Monney2009,Monney2010} 
The experimentally observed bands of $1T$-TiSe$_2$ reported 
in the ARPES measurement \cite{Cercellier2007} are well reproduced 
by BCS-like theoretical calculation assuming the excitonic insulator 
transition to be the origin of its CDW transition. \cite{Monney2009, Monney2010}
Paying attention to this similarity with $1T$-TiSe$_2$, 
the ground state of Ta$_2$NiSe$_5$ can be viewed as an excitonic 
insulator state, in which Ni $3d$-Se $4p$ valence band and Ta $5d$ 
conduction band hybridize each other due to electron-hole Coulomb 
interaction. \cite{Wakisaka2009}
In order to discuss the nature of the transition to the excitonic 
insulator state, we now analyze the temperature-dependent ARPES 
results across the transition \cite{Wakisaka2012} 
on the basis of finite temperature VCA calculations on EFKM.

Figure 4(b) shows temperature dependence in the second derivative plots
of the ARPES energy distribution curves along the Ta and Ni chain directions.
The second derivative plots at 40 K, 260 K, and 340 K are compared
with the single-particle spectral function obtained by the VCA calculations 
on EFKM for 40 K, 270 K, and 340 K which are displayed in Fig. 4(c).
The flat-top valence band at 40 K is well reproduced by 
the calculation on the EFKM with the appropriate parameter set 
as well as the previous mean-field calculation 
on the multi-band Hubbard model. \cite{Kaneko2013}
Experimentally, as the temperature increases,
the area of the flat band region in the momentum space
decreases and the band top becomes closer to the Fermi level ($E_F$).
The band dispersion deviates from the parabolic behavior 
and the flattening of the band top still remains 
even at 340 K which is above the transition temperature.
This situation is well explained by the present VCA calculation 
including the correlation effect beyond the mean-field or BCS limit.
Here, it should be noted that the flat band dispersion at 40 K exhibits 
a dip at the $\Gamma$ point which is explained by the VCA calculation 
on the q-1D EFKM.

By fitting the band positions to the band dispersion renormalized 
in a mean-field treatment assuming the excitonic insulator order 
parameter as indicated in Fig. 4, 
one can decompose the temperature dependence 
of the band top energy into the temperature dependence of $\Delta$ 
(due to the electron-hole interaction) and the temperature
dependence of the bare valence band.
The energy distribution curves (EDCs) of the $\Gamma$ point at 
40 K, 260 K, and 340 K (from which the temperature dependence 
of the bare valence band is subtracted) are compared with 
the calculated spectra for 40 K, 270 K, and 340 K in Fig. 5(a).
As the temperature increases,
the EDC peak position is shifted towards lower binding energy and 
the EDC peak width of the flat band becomes broader.
The substantial spectral broadening is reminiscent of the breaking of 
the quasi-particle peak structure, namely, the breaking of the BEC.
The broadening is similar to that observed in the pseudogap region
of the high $T_c$ cuprates although the cuprate case is more complicated
with the pseudogap assigned to a possible competing order instead of
the fluctuation of Cooper pairs. \cite{Kondo2013}
Figure 5(b) shows the temperature dependence of the order parameter
$\Delta$ and $\Delta_{\Gamma}$ (the energy position of the spectral peak 
relative to $E_F$ at the $\Gamma$ point) obtained from the calculation.
Firstly, the value of $2\Delta / k_B T_c$ is approximately 10 indicating that 
the present transition falls in the strong coupling regime.
Secondly, $\Delta_{\Gamma}$ remains finite even above the phase transition 
temperature as observed in the experiment showing that $E_G >0$ due to 
the Hartree shift by $U$.
Thirdly, the band flattening remains even above the transition temperature
indicating the excitonic fluctuation.
Interestingly, the strong coupling nature and the strong fluctuation above
the transition in Ta$_2$NiSe$_5$ are very similar to the recent ARPES observation 
on the strong coupling superconductivity with a pseudo-gap behavior 
in FeSe$_x$Te$_{1-x}$. \cite{Lubashevsky2012}
The exciton fluctuation above the phase transition temperature is 
even indicated in a semimetal $1T$-TiSe$_2$ system which is thought to 
fall in the BCS regime of the phase diagram. \cite{Monney2012a,Monney2012b} 
The present VCA calculation with fluctuation effect shows that 
the preformed exciton region in the conventional phase diagram [Fig. 1(a)]
corresponds to the pseudo-gap phase with strong excitonic fluctuation.
In addition, the transition-metal $5d$ or rare-earth $4f$ orbitals 
with spin and orbital degrees freedom under strong spin-orbit interaction
can provide a variety of spin-orbit coupled condensates derived from 
their band gap structures and spin-orbital dependent Coulomb interactions. 
Therefore, the excitonic insulators including Ta$_2$NiSe$_5$
will provide a new playground to explore physics of BCS-BEC crossover
in various bosonic systems with spin and orbital degrees freedom
which have been attracting great interest due to the new discoveries
in the ultracold atom systems. \cite{Wang2010,Lin2011,Ho2011}

\section{Conclusion}

In summary, the transition to the excitonic insulator state in Ta$_2$NiSe$_5$
has been identified as a BEC of excitons based on the comparison between 
the finite temperature VCA calculations and the ARPES results.
A flat dispersion around the top of the valence band was observed 
and assigned as the effect of excitonic coupling between 
the valence and conduction bands.
As the system exceeds the transition temperature, 
the flat feature of the valence band weakens 
though the exciton fluctuations remain finite which is 
due to the strong fluctuation effect expected 
in the BEC character of the excitonic insulator 
transition from the semiconductor phase.

\section*{Acknowledgements}

The authors would like to thank E. Hanamura, S. Koshihara, and C. Monney 
for valuable discussions.
K. Seki, Y. Wakisaka, T. Kaneko, and T. Toriyama acknowledge support from 
the JSPS Research Fellowship for Young Scientists.
This research is supported by the Grant-in-Aid for Scientific Research 
from the Japan Society for the Promotion of Science.

\clearpage
Figure captions:\\
\\
Figure 1:
(Color online)
(a) Electronic phase diagram of excitonic insulator
as a function of temperature and band gap $E_G$.
The white dashed curve indicates the boundary between 
semimetal and semiconductor regions, which roughly 
corresponds to the BCS-BEC crossover. In the BEC region,
preformed excitons are formed even above the excitonic 
insulator transition temperature.
(b) Sketch of quasi-1D crystal structure of Ta$_2$NiSe$_5$.
The excitons are formed between the electrons at the Ta chains 
and the holes at the Ni chains.
\\
\\
Figure 2:
(Color online)
Electrical resistivity and magnetic susceptibility of Ta$_2$NiSe$_5$
as functions of temperature. The arrows indicate the transition temperature.
\\
\\
Figure 3:
(Color online)
Second derivative plots of the ARPES spectra along the chain direction
of Ta$_2$NiSe$_5$ taken at various temperatures. 
The valence-band-peak positions evaluated from the energy distribution curves
are indicated by the dots with the error bars. The fit to the excitonic insulator 
band dispersion is shown by the thick solid curve.
The thin solid curve indicates the bare valence-band dispersion without 
the excitonic coupling.
\\
\\
Figure 4:
(Color online) 
(a) Second derivative plots of the ARPES spectra along the $\Gamma$-X direction taken at 40 K.
The specific flat band is formed at the top of valence band. 
(b) Band dispersions obtained from the ARPES spectra at 40 K, 260 K, and 340 K.
The dots with error bars indicate the band positions determined from the ARPES data.
The fit to the excitonic insulator band dispersion is shown by the thick solid curve.
The thin solid curve indicates the bare valence-band dispersion without the excitonic coupling.
(c) Single-particle excitation spectra for 40 K, 270 K, and 340 K obtained 
by the VCA calculation.
\\
\\
Figure 5:
(Color online) 
(a) ARPES spectra at $\Gamma$ point taken at 40 K, 260 K, and 340 K, 
and single-particle excitation spectra at $\Gamma$ point calculated 
for 40 K, 270 K, and 340 K and multiplied by the Fermi-Dirac distribution function.
(b) Temperature dependencies of the peak position of the 
flat valence band at $\Gamma$ ($\Delta_{\Gamma}$) and the order 
parameter $\Delta$, which are extracted from the ARPES results 
and are obtained from the VCA calculations.

\clearpage

\begin{figure}
  \begin{center}
      \includegraphics[width=8cm,clip]{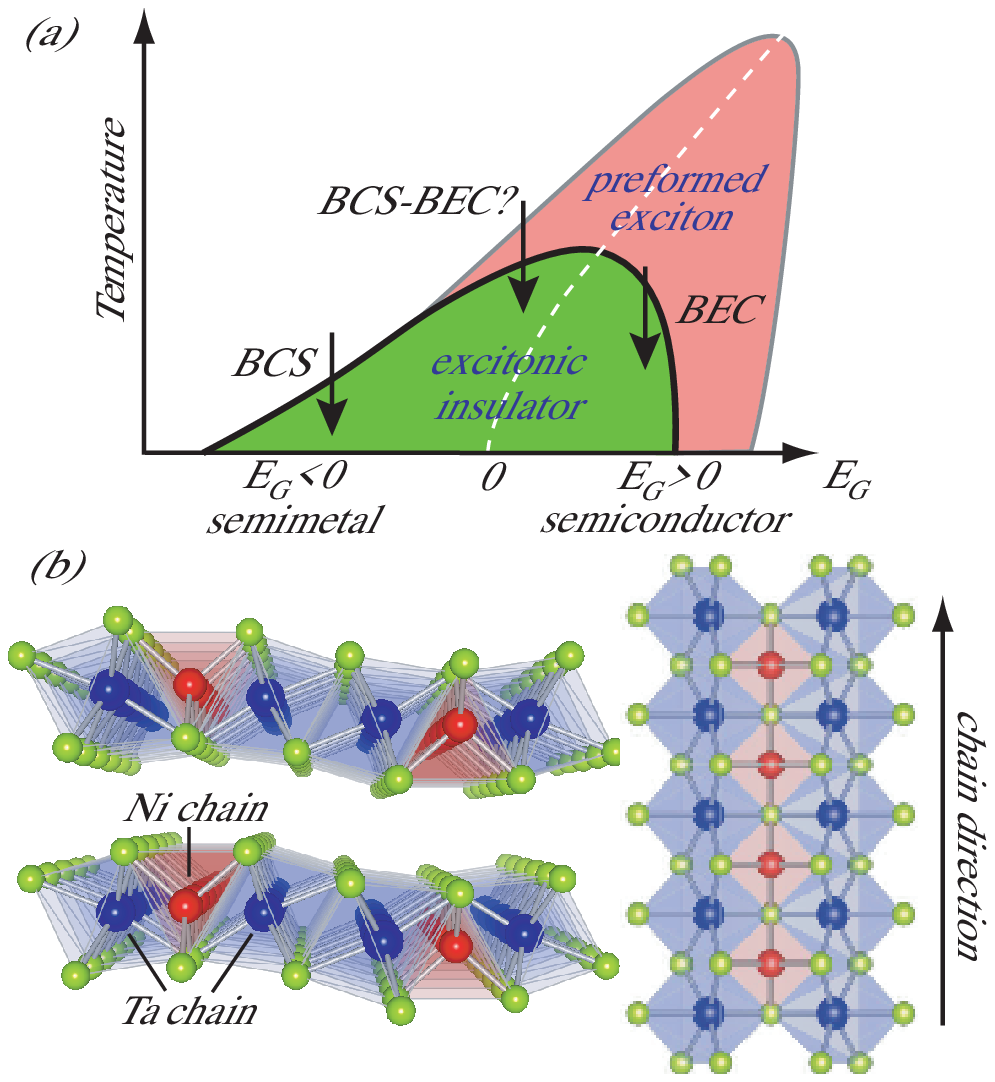}
  \end{center}
\caption{Seki et al.}
\end{figure}

\clearpage

\begin{figure}
  \begin{center}
      \includegraphics[width=8cm,clip]{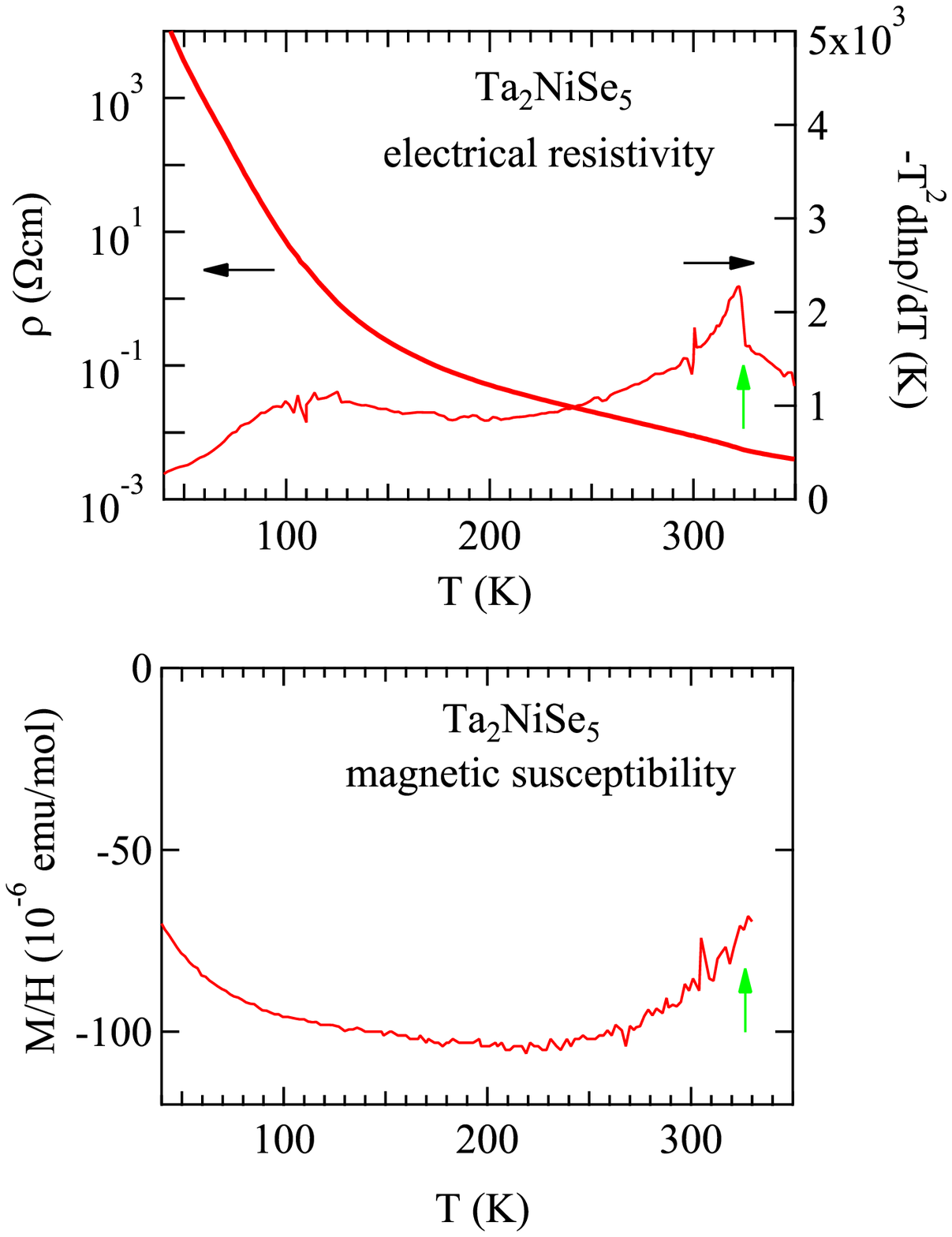}
  \end{center}
\caption{Seki et al.}
\end{figure}
\clearpage

\begin{figure}
  \begin{center}
      \includegraphics[width=10cm,clip]{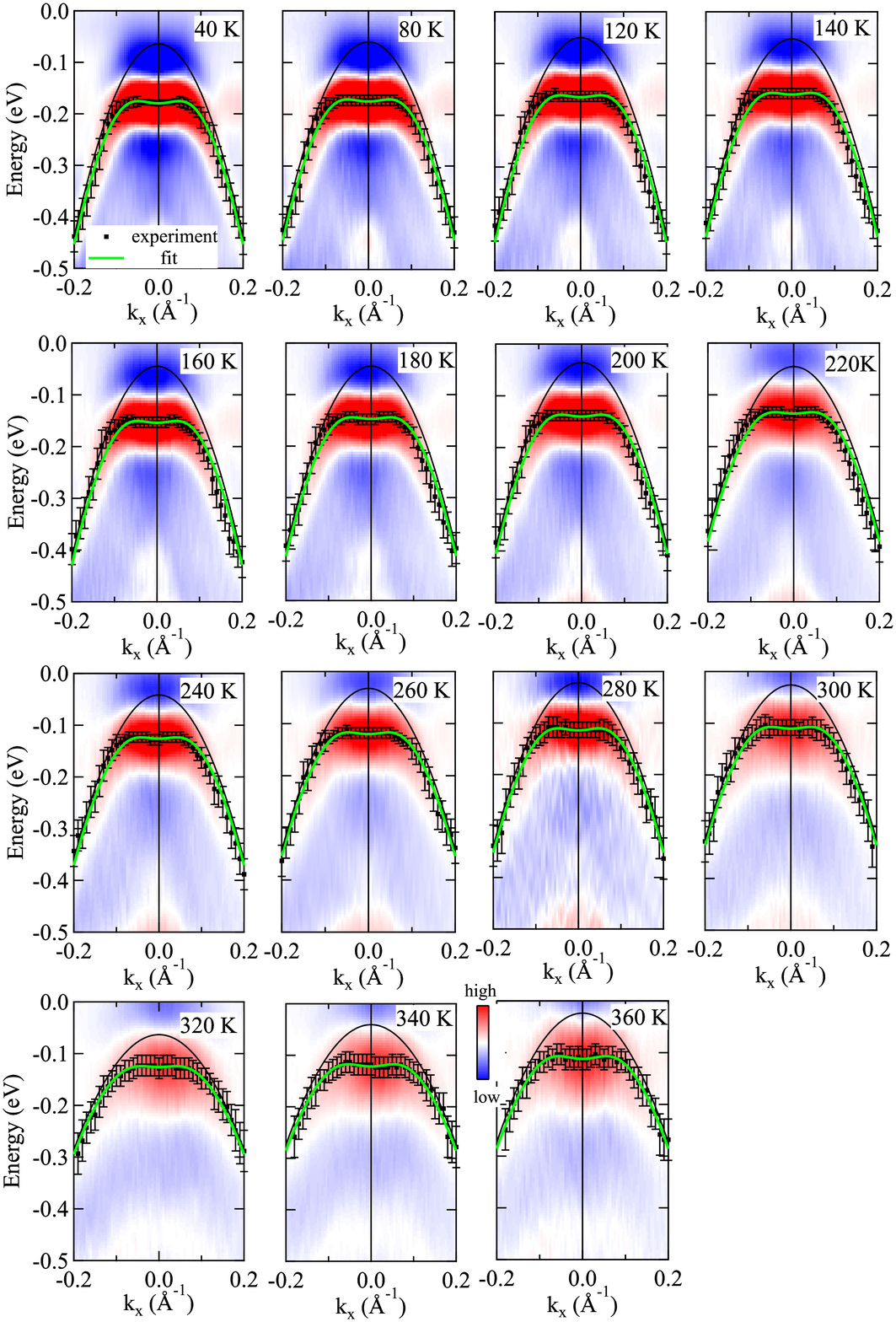}
  \end{center}
\caption{Seki et al.}
\end{figure}
\clearpage

\begin{figure}
  \begin{center}
      \includegraphics[width=10cm,clip]{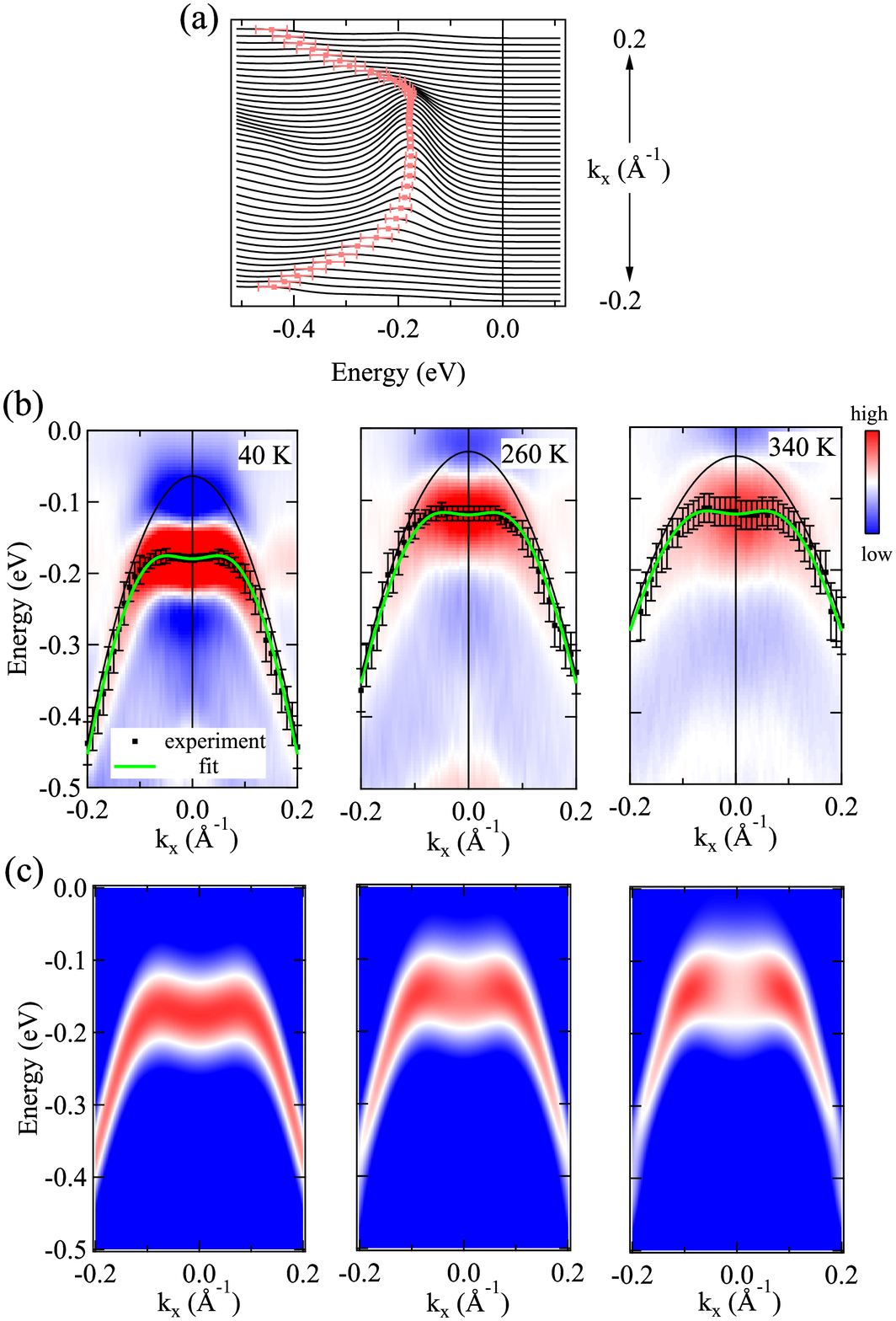}
  \end{center}
\caption{Seki et al.}
\end{figure}
\clearpage

\begin{figure}
  \begin{center}
      \includegraphics[width=8cm,clip]{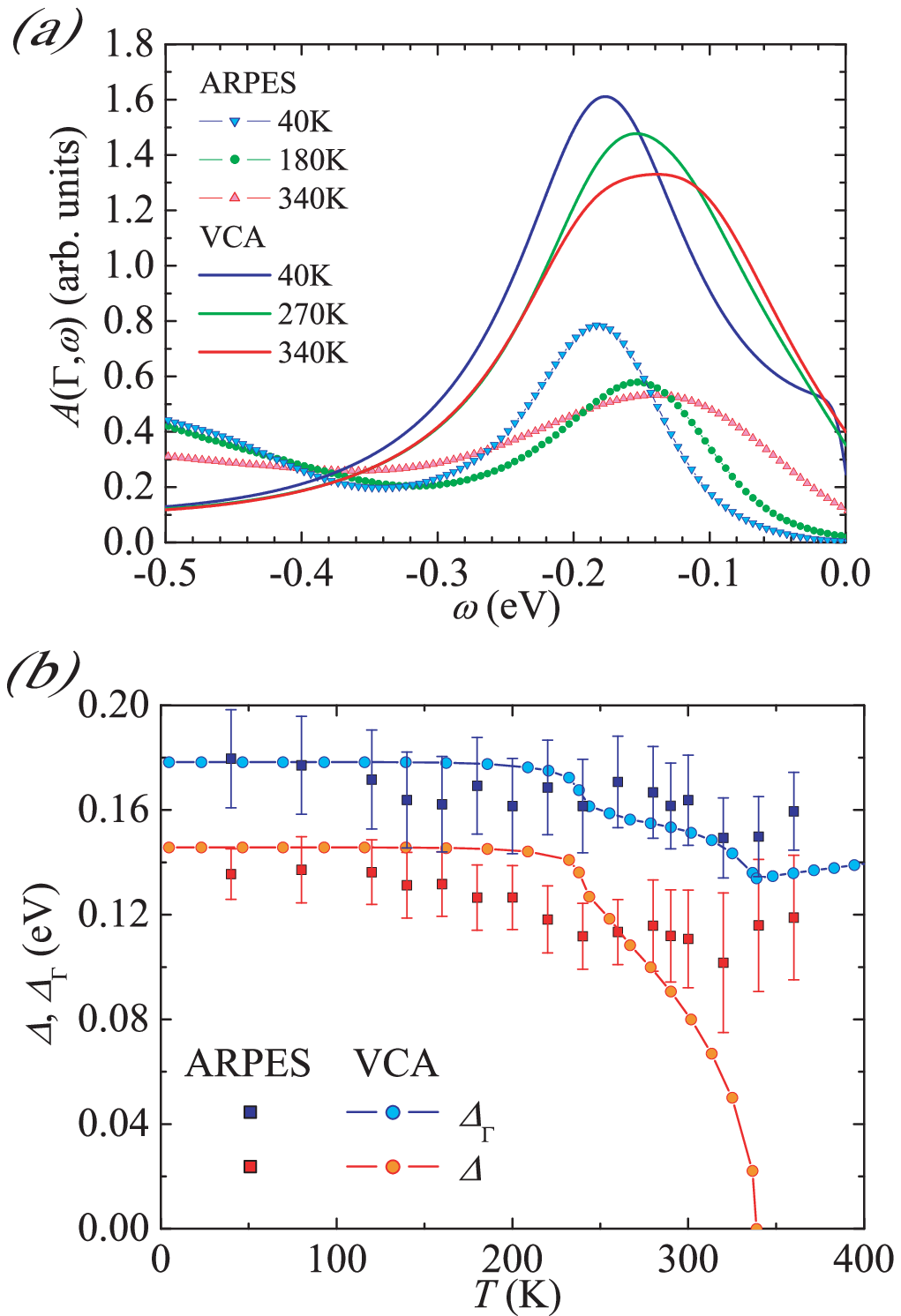}
  \end{center}
\caption{Seki et al.}
\end{figure}

\end{document}